\documentclass[twocolumn]{aastex63}
\usepackage{multirow}
\usepackage[figuresright]{rotating}
\usepackage{subfigure}
\usepackage{epic,eepic}
\usepackage{graphicx}
\usepackage{longtable}
\usepackage{float}
\usepackage{lineno}
\usepackage{pifont}
\usepackage{gensymb}
\usepackage{multirow}
\usepackage{amsmath}
\usepackage{color}

\shorttitle{}
\shortauthors{Zhao et al.}

\begin{document}

\title{Nature of HD 251108: an RS CVn binary with a long-term evolving spot}

\correspondingauthor{Song Wang}
\email{songw@bao.ac.cn}

\author{Xinlin Zhao}
\affiliation{Department of Physics and Chongqing Key Laboratory for Strongly Coupled Physics, Chongqing University, Chongqing 401331, China}

\author{Song Wang}
\affiliation{Key Laboratory of Optical Astronomy, National Astronomical Observatories, Chinese Academy of Sciences, Beijing 100101, China}
\affiliation{Institute for Frontiers in Astronomy and Astrophysics, Beijing Normal University, Beijing 102206, China}

\author{B. Fuhrmeister}
\affiliation{Th\"uringer Landessternwarte Tautenburg, Sternwarte 5, 07778 Tautenburg, Germany}
\affiliation{Hamburger Sternwarte, Universit\"at Hamburg, Gojenbergsweg 112, 21029 Hamburg, Germany}

\author{J. H. M. M. Schmitt}
\affiliation{Hamburger Sternwarte, Universit\"at Hamburg, Gojenbergsweg 112, 21029 Hamburg, Germany}

\author{Xuan Mao}
\affiliation{National Astronomical Observatories, Chinese Academy of Sciences, Beijing 100101, China}
\affiliation{School of Astronomy and Space Science, University of Chinese Academy of Sciences, Beijing 100049, China}

\author{He-Yang Liu}
\affiliation{National Astronomical Observatories, Chinese Academy of Sciences, Beijing 100101, China}

\author{Xiaohong Yang}
\affiliation{Department of Physics and Chongqing Key Laboratory for Strongly Coupled Physics, Chongqing University, Chongqing 401331, China}

\author{Jifeng Liu}
\affiliation{School of Astronomy and Space Science, University of Chinese Academy of Sciences, Beijing 100049, China}
\affiliation{New Cornerstone Science Laboratory, National Astronomical Observatories, Chinese Academy of Sciences, Beijing 100101, China}

\begin{abstract}

Recently, the Lobster Eye Imager for Astronomy (LEIA) detected the longest-lasting and most energetic stellar X-ray flare event from HD 251108. 
%
%
In this work, we re-determined the atmospheric parameters of HD 251108 using three spectroscopic observations obtained with the 2.4 m Lijiang Telescope. 
Combined with the stellar radius derived from spectral energy distribution fitting, we found that HD 251108 contains a K-type giant with a mass of approximately 1.3 $M_{\odot}$.
%
Long-term photometric monitoring over 12 years reveals a modulation suggestive of a stellar activity cycle, but inconclusive given the limited time span to date.
Light curve fitting indicates that the variations in both amplitude and shape are primarily driven by the evolution of a large spot. 
The fitting further indicates that the spot migrated from low latitudes toward the pole between 2014 and 2020, and began to recede from the pole after 2022.
Using spot parameters from light curve fitting, we found that the observed radial velocity variations arise from both the spot-induced distortions and the Keplerian orbital motion of the giant star.
Additionally, we detect a possible M-dwarf companion with a mass of approximately 0.25 $M_{\odot}$.
Our finding suggests a notable effect on the radial velocity caused by stellar magnetic activity.

\end{abstract}


\section{Introduction}
\label{intro.sec}

Stellar spots, localized cool and dark regions on stellar surfaces, are believed to be arise from local magnetic fields.
These fields suppress convective overturning, thereby inhibiting or redirecting the outward transport of energy from the stellar interior to the photosphere \citep{1938AN....264..361B,1948ZA.....25..135B}. 
As a fundamental tracer of stellar activity, the number, size, and latitude of spots are expected to evolve over a stellar cycle in response to magnetic field variations, analogous to the solar butterfly diagram that characterizes the 11-year solar cycle.
However, the characteristics and behavior of spots also vary significantly across different stellar systems. 
For instance, in some stars, spots undergo substantial evolution or migration throughout the activity cycle \citep{2013A&A...553A..40H, 2003AN....324..202R}, whereas in others they remain relatively stable over time \citep{1994A&A...281..811I,2001ASPC..223..895A}. 
In particular, polar spots in RS CVn binaries or young main-sequence stars can persist for more than a decade \citep{2009A&ARv..17..251S}. 
%

Recently, \cite{2025ApJ...980..268M} reported a superflare event on the star HD 251108, which corresponds to the X-ray source 2RXS J060415.1+124554 detected by both ROSAT \citep{2016A&A...588A.103B} and eROSITA \citep{2021A&A...647A...1P,2024A&A...682A..34M}. 
Follow-up observations revealed that the flare was extraordinarily luminous and prolonged, making it the most powerful and longest-lasting X-ray stellar flare to date.
%
%
HD 251108 also exhibits strong $H_{\alpha}$ emission and high X-ray activity during the quiescent state, reaching ${\rm log}(L_X/L_{\rm bol}) \sim -3$ \citep{2024ApJ...977....6G,2025ApJ...980..268M}. This level corresponds to the saturation regime and is unusual among giant stars \citep{2007A&A...464.1101G,2020ApJ...902..114W}. All these findings indicate strong magnetic activity in HD 251108 and motivate further investigation of this star.

Using a spot model, \cite{2024ApJ...977....6G} found that the effective temperature of HD 251108 differs by approximately 350 K between the bright and faint states.
Furthermore, \cite{2025A&A...697A.201F}, based on 149 medium-resolution spectra obtained with the Telescopio Internacional de Guanajuato Rob\'otico Espectrosc\'opico (TIGRE) telescope, detected a periodic variation in the radial velocities (RVs), implying the possible presence of a stellar companion. 
However, a periodically evolving spot can perturb the RV curve, potentially obscuring the true velocity amplitude of the orbit, and thus the nature of the secondary star.

In this work, we re-estimated the atmospheric parameters of HD 251108 using three optical spectra obtained with the 2.4 m Lijiang Telescope (LJT).
Following the method proposed by \cite{2024ApJ...963..160Z}, we employed long-term photometric LCs to examine possible decadal variations analogous to stellar activity cycles and constrain the properties of the spot.
This paper is organized as follows.
Section \ref{star.sec} describes the spectral observations and the radial velocity (RV) measurement.
Section \ref{ste_para.sec} presents stellar information of HD 251108, including the distance, atmospheric parameters, and the mass, etc.
Section \ref{lc_sc.sec} introduces the long-term LCs of this system.
In Section \ref{evo_spot.sec}, we derived the parameters of the spot by LC fitting and investigated the evolution of the spot.
Section \ref{dis_con.sec} discusses the nature of HD 251108.
Finally, We present a summary of our results in Section \ref{conclusions.sec}.

\section{Spectroscopic observation}
\label{star.sec}




We carried out high-resolution spectroscopic observations of HD 251108 using the high-resolution fiber spectrograph (HiRES) mounted on LJT at Lijiang Observatory, Yunnan Observatories, Chinese Academy of Sciences.
The Cassegrain focus of the LJT is equipped with two optical fibers. 
The first, with a diameter of 2", delivers a spectral resolution of $R \approx 32,000$ at 550 nm, while the second, with a diameter of 1.2", provides a higher resolution of $R \approx 49,000$ at the same wavelength \citep{2019RAA....19..149W}. 
HiRES covers a broad wavelength range from 320 nm to 920 nm.
Three high-resolution spectra of HD 251108 were obtained on 2022 November 10, November 30, and December 24, respectively. 
All observed wavelengths were converted to vacuum wavelengths. 
The spectral resolution of the acquired data is approximately $R \approx 32,000$ at 550 nm. 
%

Using the RV data (Table \ref{rvs.tab}) calculated from TIGRE and CARMENES spectra, \cite{2025A&A...697A.201F} confirmed that HD 251108 is a binary system composed of a K-type giant and a low-mass companion with a minimum mass of $\sim 0.1 M_{\odot}$. 
Consequently, the K-type giant contributes almost all of the total luminosity of HD 251108, even if the companion is a normal star.

\section{Stellar parameters}
\label{ste_para.sec}

\subsection{Atmospheric parameters}
\label{para.sec}

\begin{figure}
    \center
    \includegraphics[width=0.48\textwidth]{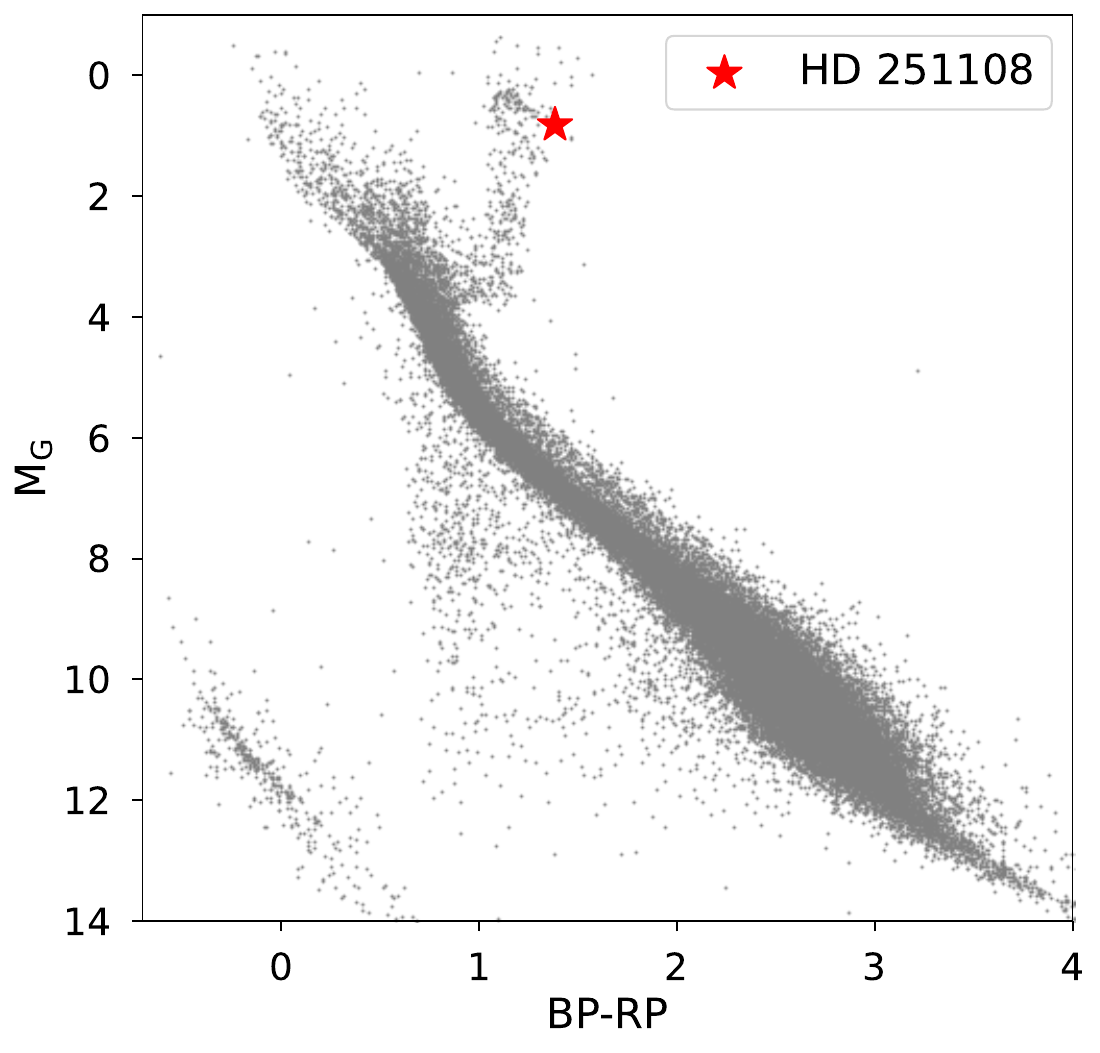}
    \caption{Position of HD 251108 on the Hertzsprung–Russell diagram. The gray points are from the Gaia DR3 with distance $d <$ 100 pc, $G_{\rm mag}$ between 4--16 mag, and galactic latitude $|b|$ $>$ 40.}
    \label{hr_diagram.fig}
\end{figure}

Gaia Data Release 3 (DR3) provides a parallax of $\varpi=1.9444\pm0.0199$ mas \citep{2021A&A...649A...1G} for HD 251108, corresponding to a distance of 
$505^{+5}_{-5}$ pc \citep{2021AJ....161..147B}.
Adopting this distance, we estimated an extinction of $A_V=0.34$ using three-dimensional maps of interstellar dust reddening \citep{2019MNRAS.483.4277C}.
Figure \ref{hr_diagram.fig} shows the location of HD 251108 in the Hertzsprung–Russell diagram, indicating that it is a nearby K-type giant star.

Previous studies have determined the atmospheric parameters of HD 251108 using both photometric and spectroscopic data. 
For instance, \cite{2019A&A...628A..94A} utilized Gaia data to report an effective temperature of $T_{\rm eff}=4545^{+406}_{-138}$ K, surface gravity of log$g=2.1^{+0.1}_{-0.1}$, and metallicity of [Fe/H]$=-0.3^{+0.3}_{-0.4}$ for HD 251108. 
Additionally, \cite{2025ApJ...984...58H} estimated these parameters by analyzing the observed XP spectrum, finding $T_{\rm}=4613\pm42$ K, log$g=2.6\pm0.1$, and [Fe/H]$=-0.2\pm0.1$.
Recently, \citet{2025A&A...697A.201F} reported a higher surface gravity of log$g=3.4\pm0.1$, derived from an analysis of TIGRE and CARMENES spectra with a template-matching method assuming solar abundances.

Using the optical spectra from LJT, we determined the atmospheric parameters of HD 251108, including the effective temperature ($T_{\rm eff}$), surface gravity (log$g$), metallicity ([Fe/H]), and projected rotational broadening velocity ($v \sin i$).
We generated a grid of synthetic stellar spectra using the Phoenix\footnote{\url{ftp://phoenix.astro.physik.uni-goettingen.de/}} model to enable interpolation in the four-dimensional parameter space of $T_{\rm eff}$, log$g$, [Fe/H], and $v \sin i$.
The synthetic spectra are sampled at a wavelength interval of 0.05 \AA. 
The grid spans $T_{\rm eff}$ from 3000 K to 6000 K in steps of 100 K, log$g$ from 1.0 to 5.0 in steps of 0.5, [Fe/H] from $-2.0$ to $+1.0$ in steps of 0.5, and $v \sin i$ from 10 km/s to 40 km/s in steps of 1 km/s.
Based on the $T_{\rm eff}$, log$g$, and [Fe/H] of each synthetic spectrum, we adopted the limb-darkening coefficients from the tables provided in \cite{2011A&A...529A..75C}.
Due to the low signal-to-noise ratio (SNR) in the blue band of the observed spectrum, we computed the $\chi^2$ values between each synthetic spectrum and the observed spectrum over the wavelength range 5500--6500\,\AA\ to construct a $\chi^2$ grid.
Using this $\chi^2$ grid, we employed a custom Markov Chain Monte Carlo (MCMC) sampler with 10 walkers and 10000 iterations to infer the stellar parameters of HD251108 (i.e., $T_{\rm eff}$, log$g$, [Fe/H], and $v \sin i$).
Table \ref{atmo_parms.tab} lists the derived stellar parameters from three independent observational epochs. 
Taking the mean of these measurements as our final estimate, we classify HD 251108 as a K-type giant with $T_{\rm eff} = 4305\pm76$ K, log$g = 2.2\pm0.2$, [Fe/H] $=-0.6\pm0.1$ , and $v \sin i = 20\pm2$ km/s, consistent with the results derived from \cite{2019A&A...628A..94A}.
The uncertainties here represent only the fitting errors from the MCMC sampling and do not include systematic errors, which likely results in an underestimated uncertainty (particularly for [Fe/H]).
Our results show a slight discrepancy with the atmospheric parameter estimates obtained by \cite{2025ApJ...984...58H}. 
This deviation may be attributed to the difference in spectral resolution: in their work, they used Gaia XP spectra to derive atmospheric parameters, whereas we employed high-resolution spectra from LJT. 
Furthermore, the machine learning method adopted by \cite{2025ApJ...984...58H} may introduce systematic biases compared to our template-matching approach, leading to the slight offset in the measurements.

\cite{2025A&A...697A.201F} mentioned that a lower surface gravity provides a better fit to some strong lines in the CARMENES spectra compared to log$g=3.44\pm0.03$.
Therefore, we also applied the same grid of synthetic spectra to measure atmospheric parameters from the averaged TIGRE and CARMENES spectra \citep{2025A&A...697A.201F}.  
Specifically, we constructed a $\chi^2$ grid using observed spectra in the wavelength range 5700--8800 \AA.
The resulting atmospheric parameters from the averaged TIGRE spectrum are $T_{\rm eff} = 4410\pm226$ K, log$g = 2.8\pm0.7$, [Fe/H] = $-0.5\pm0.3$, and $v \sin i = 18\pm3$ km/s.  
Moreover, the CARMENES spectrum yields $T_{\rm eff} = 4321\pm63$ K, log$g = 2.70\pm0.3$, [Fe/H] = $-0.5\pm0.1$, and $v \sin i = 16\pm1$ km/s.


\begin{table*}
\caption{Stellar parameters estimation for HD 251108. \label{atmo_parms.tab}}
\centering
\setlength{\tabcolsep}{5mm}
\begin{center}
\begin{tabular}{lccccc}
\hline\noalign{\smallskip}
Parameter& LJT & LJT & LJT & Averaged TIGRE & CARMENES \\
 & (2022-11-10) & (2022-11-30) & (2022-12-24) & & (2022-12-11) \\
\hline\noalign{\smallskip}
$T_{\rm eff}$ [K] & $4324\pm128$ & $4293\pm135$ & $4298\pm131$ & $4410\pm226$ & $4321\pm63$ \\
log$g$ & $2.1\pm0.4$ & $2.2\pm0.4$ & $2.2\pm0.4$ & $2.8\pm0.7$ & $2.7\pm0.3$ \\
$[\mathrm{Fe}/\mathrm{H}]$ & $-0.6\pm0.2$ & $-0.6\pm0.2$ & $-0.5\pm0.2$ & $-0.5\pm0.3$ & $-0.5\pm0.1$ \\
$v \sin i$ [km/s] & $21\pm3$ & $20\pm3$ & $18\pm3$ & $18\pm3$ & $16\pm1$ \\
\noalign{\smallskip}\hline
\end{tabular}
\end{center}
\end{table*}


\subsection{Spectral Energy Distribution Fitting}
\label{sed.sec}

\begin{figure}
    \center
    \includegraphics[width=0.48\textwidth]{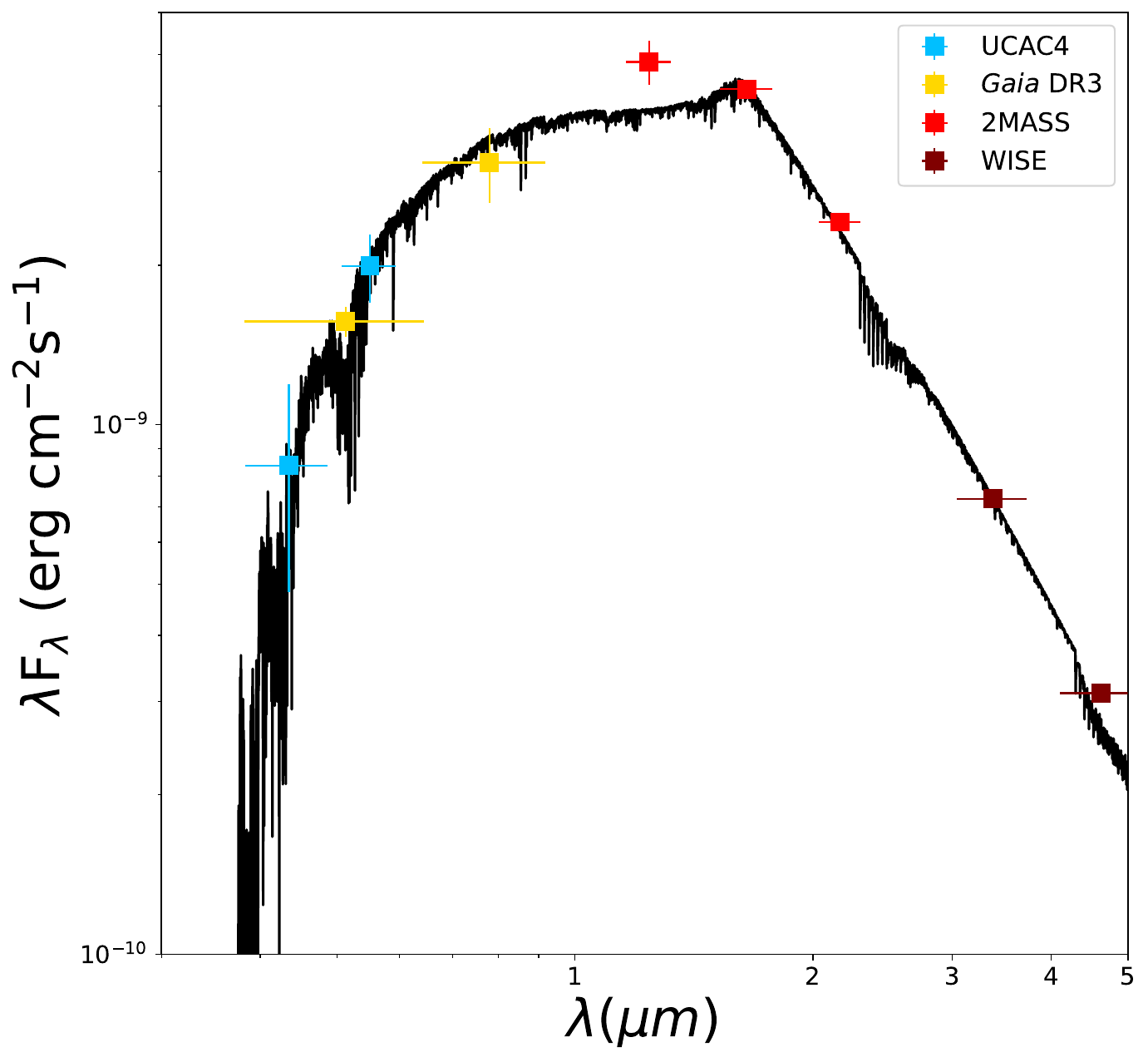}
    \caption{SED fitting of HD 251108. The black line represents the best-fit model derived by {\it SPEEDYFIT}, yielding an effective temperature of $T_{\rm eff} = 4112^{+11}_{-17}$ K and a radius of $R=14.9^{+0.2}_{-0.2} R_{\odot}$.}
    \label{sed_fitting.fig}
\end{figure}

We employed the {\it SPEEDYFIT} Python module\footnote{\url{https://speedyfit.readthedocs.io/en/stable/}} to perform a spectral energy distribution (SED) fitting in order to estimate the effective temperature and radius of HD 251108.
The SED fitting incorporated multi-band photometric data, including the $B$ and $V$ magnitudes from the fourth United States Naval Observatory CCD Astrograph Catalog (UCAC4), the $BP$, and $RP$ magnitudes from {\it Gaia}, the $J$, $H$, and $K_{\rm S}$ magnitudes from the Two Micron All Sky Survey (2MASS), and the $W1$ and $W2$ magnitudes from the Wide-field Infrared Survey Explorer ({\it WISE}).
In addition, the distance and interstellar extinction were also included as inputs during the fitting process.
The resulting effective temperature and radius from the SED fitting are $T_{\rm eff} = 4112^{+11}_{-17}$ K, and $R=14.9^{+0.2}_{-0.2} R_{\odot}$, corresponding to a luminosity of $57^{+1}_{-1} L_{\odot}$.
The best-fit SED model is shown in Figure \ref{sed_fitting.fig}.

Using the stellar radius $R$ derived from SED fitting and the surface gravity log$g$ obtained from spectroscopic analysis, we estimated the spectroscopic mass of HD 251108 through the relation $M = g R^{2} / G$.
We performed $10^{5}$ Monte Carlo sampling iterations, drawing $R = 14.9 \pm 0.2 R_{\odot}$ and log$g = 2.2\pm0.2$ from Gaussian distributions, and computed the corresponding masses. 
The final mass and its uncertainties were determined from the 50th, 16th, and 84th percentiles of the resulting mass distribution, which represent the median value and the $1\sigma$ lower and upper confidence limits, respectively.
We thus derive spectroscopic masses of $M=1.3^{+0.8}_{-0.5} M_{\odot}$ for HD 251108 from LJT spectra.
Moreover, combined with the radius $R$ from SED fitting and the $v \sin i$, we calculated a rotational inclination angle of $i=32.9^{\circ}$$^{+3.1^{\circ}}_{-3.0^{\circ}}$.

To assess the reliability of our radius and mass measurements for HD 251108, we re-estimated these parameters using stellar evolution models.
We employed the Python package {\it isochrones} \citep{2015ascl.soft03010M} to infer the evolutionary radius and mass by fitting both spectroscopic and photometric data.
The input constraints include the atmospheric parameters ($T_{\rm eff}$, $\log g$, [Fe/H]), Gaia parallax, multi-band magnitudes ($G$, $G_{\rm BP}$, $G_{\rm RP}$, $J$, $H$, $K_{\rm S}$), and extinction $A_V$, assumed as $3.1 \times E(B-V)$.
The evolutionary radius and mass inferred from {\it isochrones} are $15.0^{+0.2}_{-0.2} R_{\odot}$ and $1.0^{+0.2}_{-0.1} M_{\odot}$, respectively, in good agreement with the radius obtained from SED fitting and the spectroscopic mass derived from the LJT spectrum.
%
It should be noted that the spectroscopic masses inferred from the log$g$ measurements in the TIGRE and CARMENES spectra are about 5 $M_{\odot}$ and 4 $M_{\odot}$, substantially exceeding the mass predicted by evolutionary models.
Consequently, we adopted the stellar parameters derived from the LJT spectrum as our preferred values.

\section{Long-term light curve analysis}
\label{lc_sc.sec}

\begin{figure*}
\center 
\includegraphics[width=0.98\textwidth]{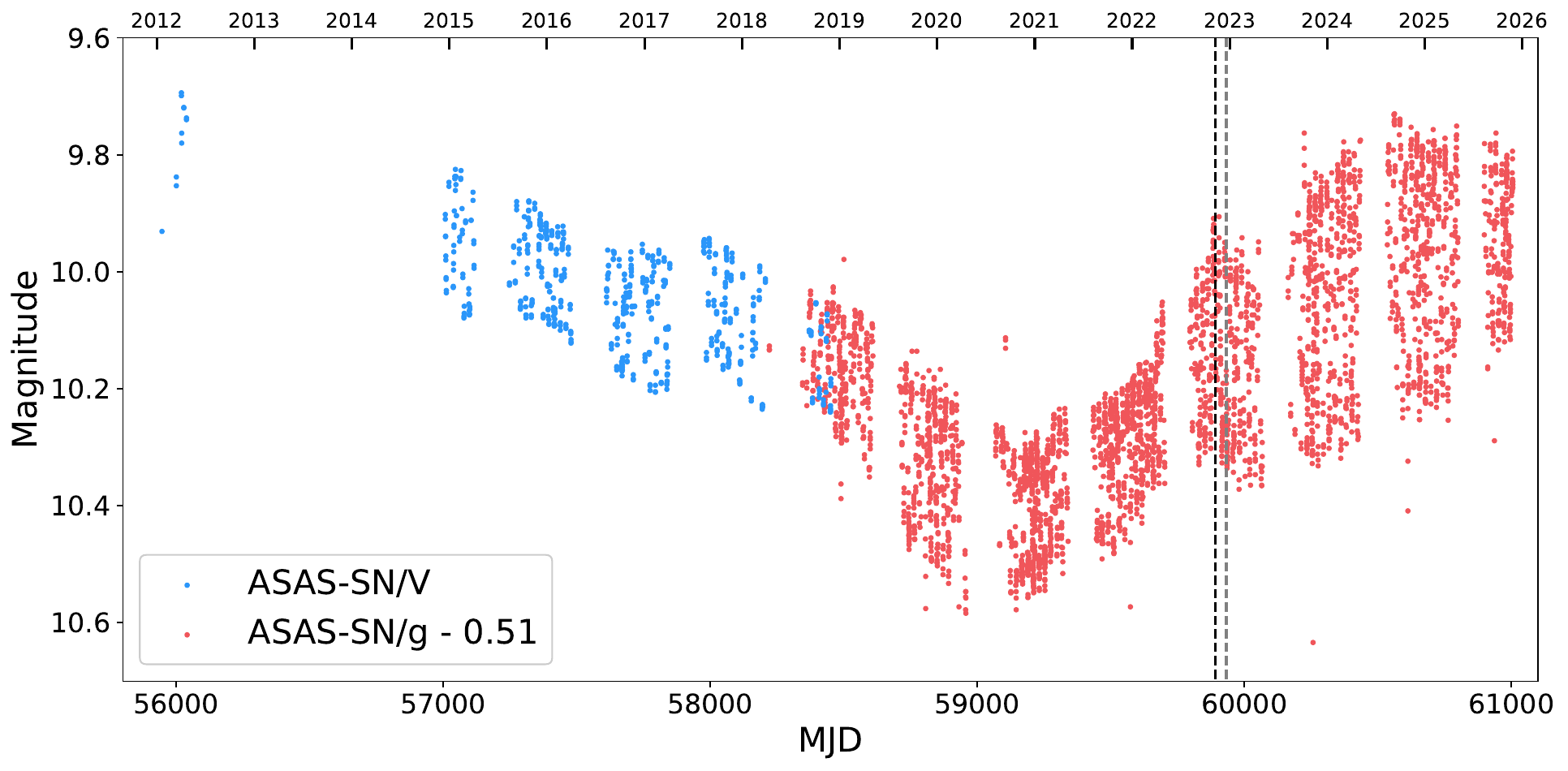}
\caption{Long-term ASAS-SN LCs of HD 251108. The black and gray dashed lines denote the peak times of the primary and secondary flares, respectively. Blue and red dots represent the $V$-band and $g$-band photometric data, respectively. All $g$ band magnitudes were shifted by -0.51 mag to align them with the $V$ band photometric scale.}
\label{asassn.fig}
\end{figure*}

HD 251108 has been monitored by ASAS-SN \citep{2014ApJ...788...48S,2017PASP..129j4502K} since 2012 (Figure \ref{asassn.fig}).
The LCs reveal a long-term variation on top of a short-term modulation owing to rotation of the spotted primary star. 
The long-term variation of about 10 years (Figure \ref{asassn.fig}) indicates the gradual evolution of large spots on the surface of the primary, suggestive of a stellar activity cycle, although the current LC span is too short to confirm a truly repetitive behavior.

\begin{figure*}
\center 
\includegraphics[width=0.98\textwidth]{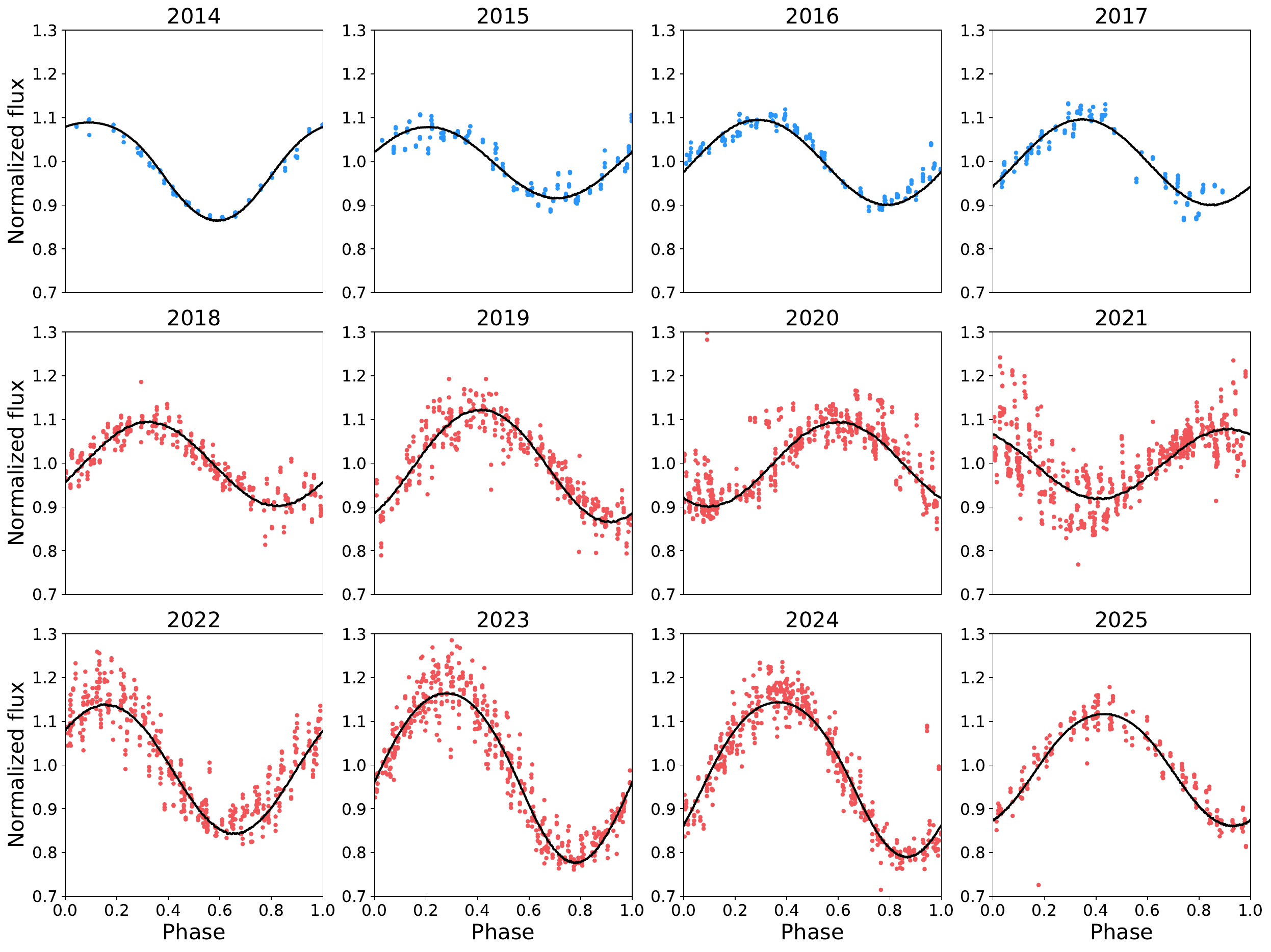}
\caption{Folded ASAS-SN LCs in $V$ (blue dots) and $g$ (red dots) bands with a period of 21.02 day. The black lines are the best-fitting models derived from the joint fitting. All $V$- and $g$-band LCs were normalized to the mean flux of their 2014 and 2018 datasets, respectively.}
\label{lcfitting.fig}
\end{figure*}

Taking the ASAS-SN LCs in the $V$ and $g$ bands, we measured the rotational period using the Lomb--Scargle (LS) method \citep{1989ApJ...338..277P}, which is specifically designed for unevenly sampled data.
Based on the previously reported period of $\approx$21.21 days from \cite{2012AcA....62...67K}, we searched for periods in the range of 15--30 days.
The LS periodogram yields a highly significant peak at $P = 21.02$ days, with a false-alarm probability $<10^{-16}$.
This period is in good agreement with the orbital period from RV measurements \citep{2025A&A...697A.201F}, indicating that HD 251108 is tidally locked.
%
Those folded LCs (Figure \ref{lcfitting.fig}) show clear, rotationally modulated flux variations.
The observed changes of their shapes possibly caused by long-term evolution of large spots on stellar surface.
%


\section{Evolution of spot properties}
\label{evo_spot.sec}

Spots, particularly polar spots, can be detected through various observational techniques such as Doppler imaging \citep{1999ApJS..121..547V,1999A&A...347..225S,2016A&A...593A.123O}, Zeeman imaging \citep{2011AN....332..866M}, and LC modeling \citep{1992A&A...259..183S,1994A&A...282..535S,1997A&A...321..811O,2002AN....323..453O,2003AN....324..202R,2024ApJ...963..160Z,2025A&A...698A.150O}.
In this section, we applied the LC fitting method to study the long-term evolution of the spot on HD 251108.

\subsection{Evolution of longitude}
\label{long.sec}

\begin{figure*}
\center 
\includegraphics[width=0.98\textwidth]{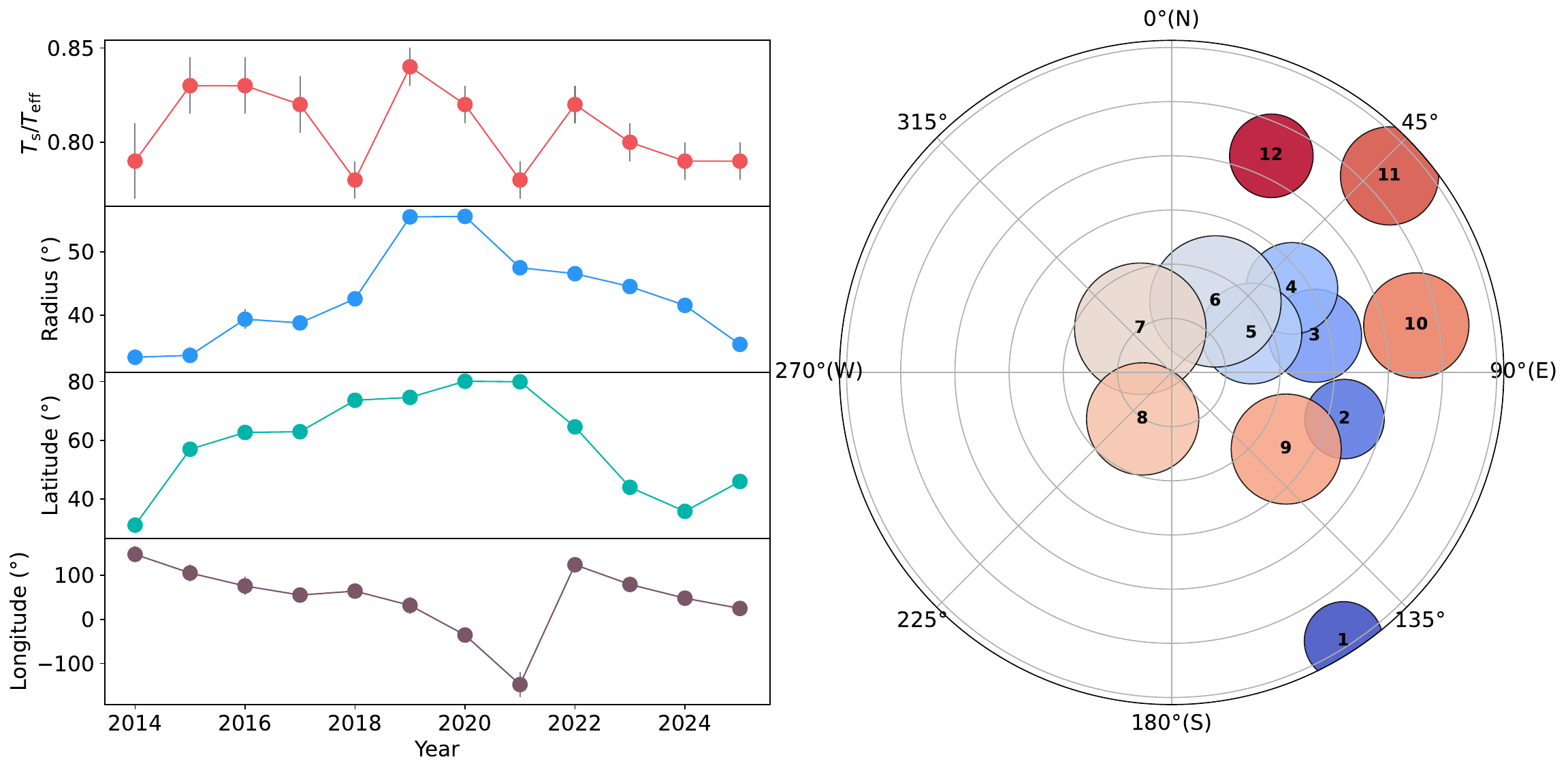}
\caption{Left panel: Evolution of the spot parameters, including temperature, angular radius, latitude, and longitude, derived from the joint fitting of ASAS-SN LCs from 2014 to 2025. The longitude values were put in the range [-180$^{\circ}$, 180$^{\circ}$] to better show their periodic behavior. Right panel: Schematic illustration of the spot evolution on the surface of HD 251108 from 2014 to 2025, shown in a polar view.}
\label{evo_spot.fig}
\end{figure*}

Compared to other spot parameters (i.e., temperature, radius, and latitude), the longitude of a spot can be directly inferred from the phase of the minimum ($\phi_{\min}$) in the observed LCs.
Thus, we first estimated the longitude of spot at each observational epoch by fitting the corresponding phase-folded LC.
Assuming a single-spot model, we fitted a sinusoidal function to the LC of each epoch to determine its $\phi_{\min}$.
For each observed LC, we produced $10^4$ mock LCs based on the photometric uncertainties.
Table \ref{spot_fitting.tab} listed the estimated longitude for each observational epoch. 
The final longitude of spot and the uncertainty were derived from the 50th, 16th, and 84th percentiles of the resulting $\phi_{\min}$ distribution.
Figure \ref{evo_spot.fig} displays the evolution of the longitude from 2014 to 2025.
Over a timespan of approximately seven years, the longitude of spot has drifted by $360^\circ$.

\subsection{LC fitting}
\label{lc_fitting.sec}

The unspotted brightness is a critical parameter in LC modeling with spots (especially polar spots) since the unspotted brightness and the spot parameters are degenerate.
A long-term joint fitting, with the unspotted brightness as a free but shared parameter for each year, can significantly break the degeneracy between the spots and the unspotted brightness of the star \citep{ 2024ApJ...963..160Z}.
Therefore, we used PHOEBE \citep{2016ApJS..227...29P, 2018ApJS..237...26H,2020ApJS..250...34C} to simultaneously fit the 12-year LCs (including ASAS-SN $V$ and $g$ LCs) by adding a single cool spots to HD 251108. 
Considering the quality of the LCs, we only used the $V$ band LCs from 2014 to 2017 and $g$ band LCs from 2018 to 2025 to perform the LC fitting.
To preserve the relative brightness across epochs, all $V$ band LCs were normalized to the mean flux of the 2014 $V$ band LC, while all $g$ band LCs were normalized to the mean flux of the 2018 $g$ band LC.
To align with the $V$ band photometric scale, all $g$ band magnitudes were shifted by $-$0.51 mag.
To improve computational efficiency in the LC fitting, we fixed the rotation period, inclination, effective temperature, radius, and mass of HD 251108 to be $P=21.02$ days, $i=32.9^{\circ}$, $T_{\rm eff}=4112$ K, $R=14.9 R_{\odot}$, and $M=1.3 M_{\odot}$, respectively.
The longitudes of the spot were also fixed based on the results presented in Sect. \ref{long.sec}.
Furthermore, to improve the computational efficiency of the LC fitting, we adopted the assumption of a single-star configuration in PHOEBE. 
This assumption remains reasonable even in a binary configuration, as HD 251108 is expected to be in synchronous rotation and the companion’s contribution to LC is negligible \citep{2025A&A...697A.201F}.

We adopted uniform priors for the spot radius and colatitude with ranges of [0, 90] and [0$^{\circ}$, 90$^{\circ}$], respectively.
Based on the analytic models from \cite{2005LRSP....2....8B}, we estimated the temperature of spot as follows:
\begin{equation}
T_{\rm s} = -895(\frac{T_{\rm eff}}{5000 \ {\rm K}})^{2} + 3755(\frac{T_{\rm eff}}{5000 \ {\rm K}}) + 808 \ {\rm K}.
\label{eq:ts}
\end{equation}
Taking the $T_{\rm eff}=4112$ K, the temperature ratio ($T_{\rm s}$/$T_{\rm eff}$) is about 0.8.
Thus, a normal prior with $\mathcal{N}(0.8, 0.1)$ was used for the temperature ratio of spot.
In the fitting, we employed an MCMC sampler with 10000 iterations.
Table \ref{spot_fitting.tab} lists the results from the best-fit model.
We obtained the unspotted magnitudes in the $V$ and $g$ bands as $9.80\pm0.01$ mag and $10.24\pm0.01$ mag, respectively.
Figure \ref{lcfitting.fig} and \ref{evo_spot.fig} show the best-fit models to the LCs and the sizes and properties of the spot. 
A large spot was identified that first migrated from low to high latitudes and then returned to low latitudes.


\begin{table*}
\caption{PHOEBE parameter estimates for the spot, including temperature ratio ($T_{\rm s}$/$T_{\rm eff}$), angular radius, colatitude, and longitude. \label{spot_fitting.tab}}
\centering
\renewcommand{\arraystretch}{1.6}
\setlength{\tabcolsep}{7mm}
\begin{center}
\begin{tabular}{lccccc}
\hline\noalign{\smallskip}
Years & $T_{\rm s}$/$T_{\rm eff}$ & Radius & Colatitude & Longitude & Band \\
 &  & ($^{\circ}$) & ($^{\circ}$) & ($^{\circ}$) &  \\
\hline\noalign{\smallskip}
2014 & $0.79^{+0.02}_{-0.02}$ & $33.3^{+1.3}_{-1.2}$ & $58.9^{+2.1}_{-1.9}$ & $147.4\pm18.8$ & V \\
2015 & $0.83^{+0.01}_{-0.02}$ & $33.6^{+1.2}_{-1.3}$ & $33.1^{+1.5}_{-2.2}$ & $105.1\pm18.9$ & V \\
2016 & $0.83^{+0.01}_{-0.02}$ & $39.3^{+1.4}_{-1.7}$ & $27.3^{+2.1}_{-1.2}$ & $75.7\pm21.2$ & V \\
2017 & $0.82^{+0.01}_{-0.02}$ & $38.8^{+1.1}_{-1.1}$ & $27.1^{+2.3}_{-1.6}$ & $55.0\pm16.3$ & V \\
2018 & $0.78^{+0.01}_{-0.01}$ & $42.6^{+0.3}_{-0.4}$ & $16.4^{+0.5}_{-0.4}$ & $64.0\pm13.2$ & g \\
2019 & $0.84^{+0.01}_{-0.01}$ & $55.5^{+0.5}_{-0.4}$ & $15.4^{+0.4}_{-0.4}$ & $31.7\pm19.3$ & g \\
2020 & $0.82^{+0.01}_{-0.01}$ & $55.6^{+0.3}_{-0.5}$ & $9.9^{+0.4}_{-0.4}$ &  $324.4\pm10.4$ & g \\
2021 & $0.78^{+0.01}_{-0.01}$ & $47.5^{+0.3}_{-0.3}$ & $10.1^{+0.4}_{-0.4}$ & $212.0\pm28.3$ & g \\
2022 & $0.82^{+0.01}_{-0.01}$ & $46.6^{+0.4}_{-0.5}$ & $25.4^{+0.4}_{-0.4}$ & $123.8\pm12.7$ & g \\
2023 & $0.80^{+0.01}_{-0.01}$ & $44.5^{+0.4}_{-0.3}$ & $46.0^{+0.4}_{-0.4}$ & $79.1\pm5.4$ & g \\
2024 & $0.79^{+0.01}_{-0.01}$ & $41.5^{+0.4}_{-0.3}$ & $54.2^{+0.6}_{-0.4}$ & $47.9\pm5.7$ & g \\
2025 & $0.79^{+0.01}_{-0.01}$ & $35.4^{+0.3}_{-0.3}$ & $44.0^{+0.4}_{-0.4}$ & $24.7\pm13.7$ & g \\
\noalign{\smallskip}\hline
\end{tabular}
\end{center}
\end{table*}

\subsection{Evolution of the spot}
\label{evo.sec}

Based on the results of LC fitting, Figure \ref{evo_spot.fig} illustrates the evolution of the spot on the stellar surface between 2014 and 2025.
The migration of the spot can be divided into two distinct phases: an initial poleward migration from low to high latitudes, followed by an equatorward return.
During the first phase (2014–2021), the spot gradually migrated from a low latitude of approximately $30^{\circ}$ to near-polar latitudes ($\sim 80^{\circ}$). 
Over this period, it completed nearly one full cycle in longitude while its radius steadily increased. 
Upon reaching the polar region, the spot attained its maximum radius of $\sim 56^{\circ}$ and remained stable near the pole for about two years.
In the second phase (2022-2025), the spot began migrating equatorward, accompanied by a gradual decrease in its radius. 
In 2022, the colatitude of the spot closely matched the stellar inclination angle of HD 251108, indicating it was nearly aligned with our line of sight. 
Notably, the spot still had a large radius of approximately $47^{\circ}$ at that time.
It's interesting that the largest projected area of the spot on the visible hemisphere occurred in 2022, the year that the super X-ray flare was detected.

\section{Discussion}
\label{dis_con.sec}

\subsection{Other systems with large spots}
\label{compri.sec}

There are many stars known to host large spots \citep{2002AN....323..453O,2003AN....324..202R,2010AN....331..794O,2012AN....333..138O,2024ApJ...963..160Z}.
Long-term photometric monitoring revealed that UZ Librae \citep{2002AN....323..453O}, IM Pegasi \citep{2003AN....324..202R}, and TIC 16320250 \citep{2024ApJ...963..160Z} host a large, evolving polar spot.
%

\cite{2002AN....323..453O} analyzed the evolution of the spots on UZ Librae using nine years of photometric data and identified a spot cycle with a period of approximately 4.8 years.
In their LC modeling, they found that a configuration consisting of two equatorial spots and one polar spot was required to adequately reproduce the observed photometric variations.
Throughout the nine-year dataset, the sizes and longitudes of the two equatorial spots remained remarkably stable.
In contrast, the polar spot maintained a stable radius between 20$^{\circ}$ and 40$^{\circ}$, while its longitude drifted by approximately 260$^{\circ}$.

\cite{2003AN....324..202R} analyzed 23 years of photometric data of IM Pegasi using a two-spot model.
The LC modeling revealed the presence of a polar spot and a second spot located at a latitude of approximately $-50^{\circ}$.
Both spots exhibit large angular radii, varying between 40$^{\circ}$ and 60$^{\circ}$. 
Notably, the polar spot shows clear cyclic variations in both its radius and longitude, with periods of 29.8 years and 10.4 years, respectively.

\cite{2024ApJ...963..160Z} used 20 years of photometric data to reveal a spot cycle of approximately 10 years in TIC 16320250.
They identified two polar spots with angular radii varying between 20$^{\circ}$ and 40$^{\circ}$.
Similar to IM Pegasi, both polar spots of TIC 16320250 exhibited significant evolution in radius and longitude.
For the one spot, the radius decreased as its latitude decreased, indicating a migration toward the equator.


Compared with the systems discussed above, the radii derived from LC fitting fall within reasonable ranges.
Moreover, the radius of the spot increased significantly as it migrated toward the pole, and decreased markedly as it moved away from the polar region toward lower latitudes.
This evolutionary behavior is consistent with the spot characteristics observed in the systems described above.
%

\begin{figure*}
\center 
\includegraphics[width=0.98\textwidth]{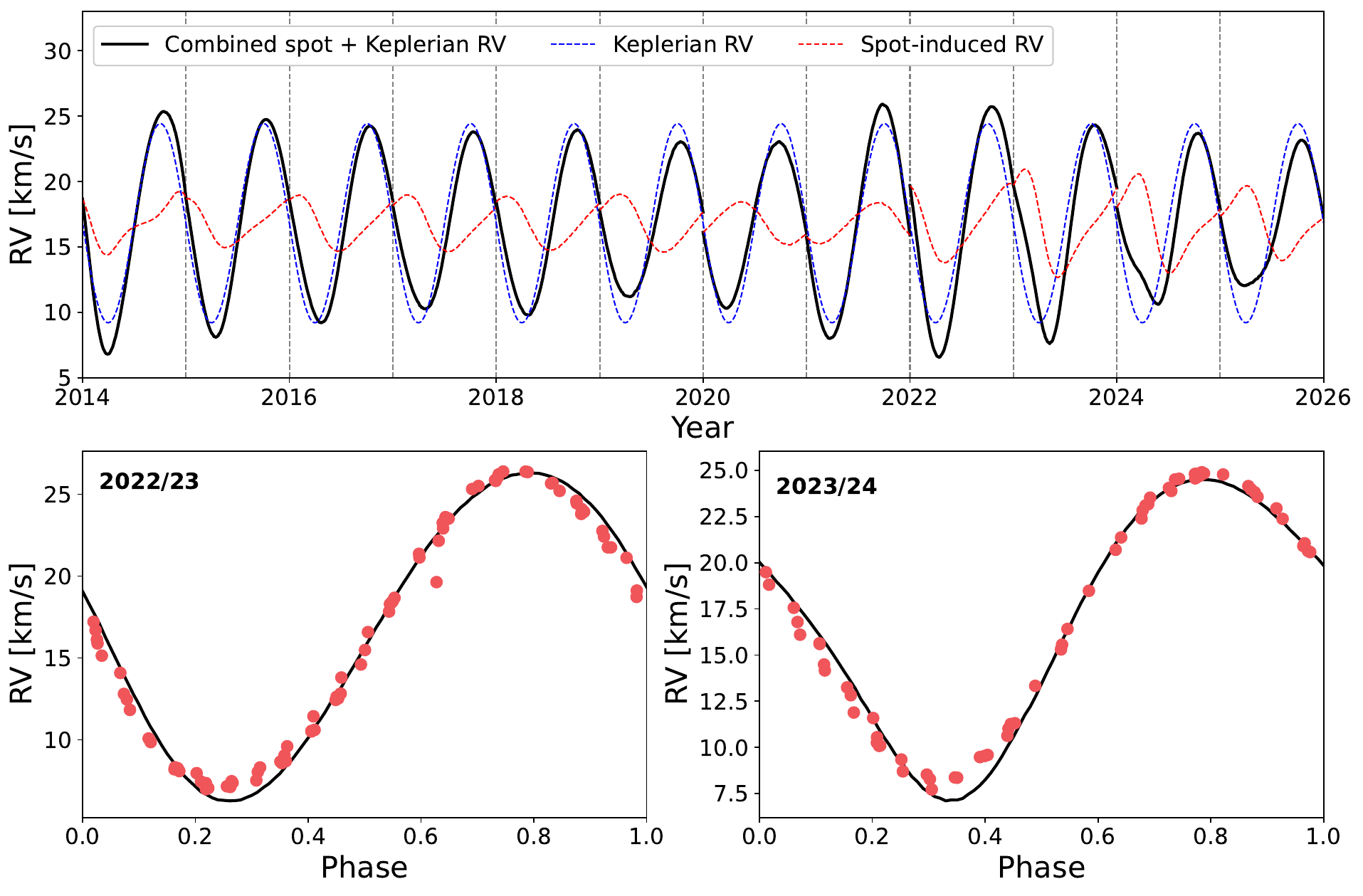}
\caption{Theoretical RV curves generated with PHOEBE, assuming a companion mass of 0.25 $M_{\odot}$. Top panel: Theoretical RV curves from 2014 to 2025. These RV curves for each year are phase-folded using the period of 21.02 days and plotted cumulatively. The blue line shows the Keplerian RV due to orbital motion, the red line represents the spot-induced RV variation, and the black line is the combined  RV curve. Bottom panel: Comparison of the observed RV measurements (red dots) from the TIGRE telescope with the theoretical RV curve (black solid line).}
\label{rvs.fig}
\end{figure*}

\subsection{Nature of HD 251108}
\label{nature.sec}

\cite{2025A&A...697A.201F} measured the RVs of HD 251108 using multiple high-resolution spectra, revealing the periodic RV variations with a period of 21.1 days and a RV semi-amplitude of $K=8.980\pm0.028$ km/s (Table \ref{rvs.tab}).
When folding the RV data on this period, they found that the amplitude decreases from the 2022/2023 to the 2023/2024 observing seasons, and the phases of the maximum and minimum RV appear to shift slightly toward higher or lower values.
In their subsequent analysis, they concluded that the observed RV signal likely arises from the combined effects of a spot and a low-mass (M-dwarf) binary.

In this work, we investigated the origin of the RV variations and the nature of the companion using the results in LC fitting (Sect. \ref{lc_fitting.sec}).
First, we adopted a single-star configuration and used PHOEBE to generate theoretical RV curves based on the best-fit parameters listed in Table \ref{spot_fitting.tab}. 
Figure \ref{rvs.fig} shows the resulting RV curves corresponding to the observing epochs from 2014 to 2025.
Notably, the theoretical amplitudes of the spot-induced RV curves (the red dashed line in Figure \ref{rvs.fig}) increase from 2022/23 to 2023/24, in contrast to the declining amplitude of RV data reported by \cite{2025A&A...697A.201F}.
This discrepancy suggests that the observed RV variations cannot be fully attributed to the spot alone.
%
%
Consequently, the temporal evolution of the RV data is likely driven by the migration or evolution of the spot across the stellar surface.

Second, we considered the scenario of a synchronously rotating binary system.
Given the negligible impact of the companion on the LC, we kept the LC solution from Sect. \ref{lc_fitting.sec} unchanged when modeling the system as a binary in PHOEBE.
We used a grid ranging from 0.1 $M_{\odot}$ to 0.4 $M_{\odot}$ in steps of 0.01 $M_{\odot}$ for the mass of the companion.
Finally, we find that when the phase of the spot-induced RV is offset by 0.5 (i.e., in anti-phase) relative to that of the orbital RV signal, a companion with a mass of approximately 0.25 $M_{\odot}$ successfully reproduces both the amplitude and temporal evolution of the RV variations reported by \cite{2025A&A...697A.201F} (Figure \ref{rvs.fig}).

%

Although giant RS CVn systems can host subgiant companions, they typically have main-sequence secondaries \citep{1993A&AS..100..173S}. 
Given evolutionary time-scale considerations, a subgiant companion to HD 251108 would be expected to have a mass just below that of the primary ($\sim 1 M_{\odot}$), which is significantly higher than our estimated value of $\sim 0.25 M_{\odot}$.
Thus, we confirm that HD 251108 is a giant RS CVn with a low-mass companion of approximately $0.25 M_{\odot}$, probably an M dwarf.
The observed RV variations are primarily driven by the migration and evolution of a spot on the surface of the K-type giant star.


\section{Conclusions}
\label{conclusions.sec}

We re-determined the atmospheric parameters of HD 251108 using optical spectra obtained with LJT.
A grid of PHOENIX model was constructed, spanning a range of effective temperatures ($T_{\rm eff}$), surface gravities (log$g$), metallicities ([Fe/H]), and projected rotational velocities ($v \sin i$).  
The best-fit atmospheric parameters were found to be $T_{\rm eff} = 4305\pm76$ K, log$g = 2.2\pm0.2$, [Fe/H] $=-0.6\pm0.1$ , and $v \sin i = 20\pm2$ km/s, in agreement with previous studies \citep{2019A&A...628A..94A,2025ApJ...984...58H}.  
SED fitting yielded a stellar radius of $14.9^{+0.2}_{-0.2} R_{\odot}$ and a spectroscopic mass of $1.3^{+0.8}_{-0.5} M_{\odot}$, consistent with predictions from single-star evolutionary models.

Over a 12-year baseline, the ASAS-SN LCs exhibited significant variations in both amplitude and shape, suggesting the possible evolution or migration of a spot on stellar surface.  
A LS periodogram analysis of the ASAS-SN LCs returned a dominant period of approximately 21.02 days.
This period is consistent with the orbital period derived from RV measurements, suggesting the rotation of the giant star and orbital motion are locked.
The contribution of the companion to the LCs was negligible, as the mass function indicated a low-mass secondary.  
%
The conspicuous long-term photometric variation is at least a decade, and suggestive of an activity cycle; but whether the modulation is cyclic or stochastic is not settled owing to the short span of the existing LC.

In addition, a joint fitting was applied to multi-year LCs to estimate the parameters of the spot, including temperature, longitude, latitude, and angular radius, over the interval from 2014 to 2025.
The analysis revealed periodic variations in the longitude, latitude, and angular radius.
Between 2014 and 2021, the spot migrated poleward from low latitudes while its angular radius steadily increased.
When the spot approached polar regions (approximately 2018 to 2021), its angular radius reached a maximum value of about $\sim 56^{\circ}$.
Starting around 2022, the spot began migrating equatorward from the polar regions, accompanied by a gradual decrease in angular radius.
During this period, the spot drifted in longitude by approximately one and a half rotations, or about $483^{\circ}$.
The observed evolution of the spot on HD 251108 closely resembles the characteristic behavior of polar spots, supporting the reliability of the joint-fit results.
This large spot may also help explain its strong $H_{\alpha}$ emission and high X-ray activity level.

Finally, using the parameters derived from the joint fitting, synthetic RV curves for HD 251108 over the 2014--2025 were generated with the PHOEBE code. 
These models were constrained by RV measurements obtained in 2022 and 2023.
%
The analysis revealed that spot-induced RV semi-amplitudes ranged from approximately 2 km/s to 4 km/s, exceeding the estimate reported by \cite{2025A&A...697A.201F}.  
Since the periodic evolution of the spot, the RV semi-amplitude exhibited a corresponding periodic modulation between 2014 and 2025.  
Furthermore, a companion mass of approximately $0.25 M_{\odot}$ was determined by fitting the RV data from 2022/23 and 2023/24.

\begin{acknowledgements}

We thank the anonymous referee for very helpful comments and suggestions that significantly improved paper.
This work was funded by the Strategic Priority Program of the Chinese Academy of Sciences under grant number XDB1160302 (Song Wang), the National Key Research and Development Program of China under grant number 2023YFA1607901 (Song Wang), the National Natural Science Foundation of China (NSFC) under grant number 12588202 (Jifeng Liu), the NSFC under grant number 12273057 (Song Wang), science research grants from the China Manned Space Project (Song Wang), the New Cornerstone Science Foundation through the New Cornerstone Investigator Program and the XPLORER PRIZE (Jifeng Liu), Chongqing Natural Science Foundation under grant number CSTB2023NSCQ-MSX0093 (Xiaohong Yang), the NSFC under grand number 12547101 (Xinlin Zhao), and the Postdoctoral Fellowship Program of China Postdoctoral Science (CPSF) under grant number GZB20250739 (Xinlin Zhao).

\end{acknowledgements}

\bibliographystyle{aasjournal}
\bibliography{main.bib}{}   

\clearpage
\appendix
\renewcommand*\thetable{\Alph{section}.\arabic{table}}
\renewcommand*\thefigure{\Alph{section}\arabic{figure}}

\section{Radial velocity measurements}
\label{rvdata.sec}

Here, we present the RV measurements of HD 251108 derived from the TIGRE telescope in Table \ref{rvs.tab}.

\begin{table*}
\caption{RV measurements for HD 251108 from the TIGRE telescope. \label{rvs.tab}}
\centering
\renewcommand{\arraystretch}{1}
\setlength{\tabcolsep}{2mm}
\begin{center}
\begin{tabular}{cccc|cccc}
\hline\noalign{\smallskip}
BJD & RV & Uncertainty & SNR & BJD & RV & Uncertainty & SNR \\
(day) & (km/s) & (km/s)  & & (day) & (km/s) & (km/s) & \\
\hline\noalign{\smallskip}
2459895.80302 & 8.10 & 0.23 & 55.9 & 2459977.69901 & 14.09 & 0.22 & 88.5 \\
2459896.79183 & 7.36 & 0.24 & 50.4 & 2459980.70399 & 7.46 & 0.21 & 84.4 \\
2459897.78667 & 7.39 & 0.27 & 47.1 & 2459983.66397 & 8.67 & 0.21 & 87.3 \\
2459898.81171 & 8.32 & 0.23 & 74.0 & 2459986.66573 & 14.60 & 0.19 & 78.0 \\
2459899.82282 & 9.61 & 0.25 & 72.1 & 2460031.61159 & 22.16 & 0.19 & 59.7 \\
2459900.80029 & 11.44 & 0.27 & 65.7 & 2460038.61254 & 21.12 & 0.21 & 70.1 \\
2459901.84205 & 13.81 & 0.34 & 23.6 & 2460043.61327 & 7.97 & 0.20 & 64.0 \\
2459902.83335 & 16.58 & 0.16 & 72.5 & 2460262.79128 & 19.63 & 0.16 & 70.0 \\
2459906.95029 & 25.50 & 0.23 & 59.2 & 2460264.78839 & 23.52 & 0.22 & 70.0 \\
2459907.87209 & 26.38 & 0.23 & 38.1 & 2460266.79369 & 24.84 & 0.24 & 67.6 \\
2459908.78295 & 26.35 & 0.26 & 68.2 & 2460268.79118 & 23.56 & 0.24 & 78.0 \\
2459909.97477 & 25.20 & 0.29 & 35.7 & 2460272.76649 & 16.10 & 0.17 & 67.2 \\
2459910.78098 & 23.80 & 0.23 & 67.2 & 2460274.76812 & 11.89 & 0.21 & 68.5 \\
2459911.77191 & 21.76 & 0.25 & 66.8 & 2460280.77237 & 11.31 & 0.16 & 22.5 \\
2459912.84397 & 18.73 & 0.25 & 76.0 & 2460286.89126 & 24.54 & 0.22 & 82.9 \\
2459913.77890 & 15.89 & 0.24 & 46.3 & 2460290.74830 & 22.38 & 0.21 & 66.2 \\
2459913.93766 & 15.14 & 0.24 & 61.9 & 2460294.71286 & 14.17 & 0.23 & 54.0 \\
2459914.98646 & 11.83 & 0.21 & 19.0 & 2460298.69656 & 7.72 & 0.22 & 66.8 \\
2459915.75974 & 9.87 & 0.19 & 53.7 & 2460300.77955 & 9.59 & 0.23 & 56.3 \\
2459916.76451 & 8.27 & 0.19 & 60.9 & 2460306.66264 & 23.10 & 0.19 & 61.5 \\
2459918.77592 & 7.48 & 0.23 & 64.6 & 2460306.76321 & 23.17 & 0.18 & 77.6 \\
2459919.75143 & 8.04 & 0.17 & 68.4 & 2460308.75942 & 24.90 & 0.19 & 72.0 \\
2459920.74647 & 9.05 & 0.17 & 66.2 & 2460310.73220 & 23.81 & 0.21 & 40.8 \\
2459921.75113 & 10.52 & 0.20 & 62.8 & 2460310.81323 & 16.15 & 0.58 & 14.7 \\
2459922.74572 & 12.53 & 0.23 & 69.1 & 2460312.79942 & 20.57 & 0.20 & 57.4 \\
2459923.74631 & 15.49 & 0.24 & 69.3 & 2460313.65754 & 18.81 & 0.20 & 71.3 \\
2459924.77549 & 18.46 & 0.21 & 66.2 & 2460315.70719 & 14.49 & 0.21 & 68.3 \\
2459925.76015 & 21.36 & 0.21 & 69.1 & 2460317.67539 & 10.26 & 0.23 & 28.9 \\
2459926.74516 & 23.60 & 0.21 & 67.0 & 2460317.75758 & 10.09 & 0.28 & 31.6 \\
2459927.73903 & 25.32 & 0.22 & 75.8 & 2460317.79862 & 10.10 & 0.22 & 31.3 \\
2459928.72934 & 26.21 & 0.23 & 70.4 & 2460318.64569 & 8.70 & 0.21 & 50.2 \\
2459929.72931 & 26.38 & 0.21 & 74.6 & 2460320.66799 & 8.36 & 0.22 & 49.5 \\
2459930.73994 & 25.69 & 0.23 & 66.9 & 2460322.66495 & 11.27 & 0.21 & 68.9 \\
2459931.82361 & 24.09 & 0.21 & 44.7 & 2460324.77786 & 16.41 & 0.20 & 65.5 \\
2459931.91049 & 23.93 & 0.32 & 17.3 & 2460326.77832 & 21.37 & 0.19 & 58.2 \\
2459932.91408 & 21.75 & 0.24 & 63.8 & 2460328.78894 & 24.51 & 0.29 & 26.7 \\
2459933.87542 & 19.12 & 0.21 & 98.0 & 2460329.71323 & 24.71 & 0.28 & 16.9 \\
2459934.77800 & 16.13 & 0.24 & 73.5 & 2460330.57848 & 24.77 & 0.20 & 31.1 \\
\noalign{\smallskip}\hline
\end{tabular}
\end{center}
\end{table*}

\begin{table*}
\caption{Continued.}
\centering
\renewcommand{\arraystretch}{1}
\setlength{\tabcolsep}{2mm}
\begin{center}
\begin{tabular}{cccc|cccc}
\hline\noalign{\smallskip}
BMJD & RV & Uncertainty & SNR & BMJD & RV & Uncertainty & SNR \\
(day) & (km/s) & (km/s)  & & (day) & (km/s) & (km/s) & \\
\hline\noalign{\smallskip}
2459935.78616 & 12.81 & 0.22 & 27.9 & 2460333.56609 & 20.90 & 0.25 & 29.8 \\
2459935.89573 & 12.47 & 0.22 & 64.9 & 2460333.74393 & 20.63 & 0.20 & 73.4 \\
2459936.70751 & 10.09 & 0.20 & 81.2 & 2460335.74056 & 16.79 & 0.20 & 70.1 \\
2459937.66892 & 8.20 & 0.24 & 29.5 & 2460337.72657 & 12.82 & 0.20 & 60.7 \\
2459938.84351 & 6.99 & 0.21 & 31.9 & 2460338.70558 & 10.55 & 0.24 & 46.0 \\
2459938.92876 & 7.06 & 0.27 & 27.6 & 2460340.67439 & 8.28 & 0.21 & 20.2 \\
2459940.72011 & 7.53 & 0.22 & 74.8 & 2460342.71055 & 9.54 & 0.19 & 50.6 \\
2459941.67197 & 8.59 & 0.20 & 65.6 & 2460344.60055 & 13.33 & 0.21 & 43.6 \\
2459941.82136 & 8.71 & 0.23 & 82.5 & 2460346.60123 & 18.48 & 0.19 & 63.6 \\
2459942.77700 & 10.55 & 0.21 & 47.3 & 2460348.60172 & 22.83 & 0.19 & 77.8 \\
2459942.89075 & 10.61 & 0.20 & 67.7 & 2460350.60236 & 24.81 & 0.17 & 77.0 \\
2459943.67756 & 12.44 & 0.18 & 71.5 & 2460358.61305 & 13.25 & 0.22 & 68.3 \\
2459943.86243 & 12.83 & 0.18 & 71.8 & 2460360.63393 & 9.35 & 0.21 & 68.8 \\
2459945.70154 & 18.29 & 0.24 & 33.7 & 2460362.63049 & 8.38 & 0.22 & 69.5 \\
2459945.76152 & 18.39 & 0.19 & 35.6 & 2460364.62698 & 11.02 & 0.20 & 64.4 \\
2459945.86992 & 18.67 & 0.20 & 48.9 & 2460366.62366 & 15.56 & 0.17 & 65.3 \\
2459946.79826 & 21.13 & 0.21 & 72.4 & 2460368.62009 & 20.70 & 0.18 & 72.3 \\
2459947.66645 & 23.25 & 0.21 & 64.4 & 2460370.60624 & 24.04 & 0.21 & 35.5 \\
2459947.75080 & 23.40 & 0.22 & 27.9 & 2460370.69558 & 23.88 & 0.18 & 27.2 \\
2459947.88601 & 23.51 & 0.23 & 41.7 & 2460371.57560 & 24.80 & 0.34 & 16.2 \\
2459949.61605 & 25.86 & 0.22 & 40.9 & 2460371.66565 & 24.64 & 0.18 & 56.2 \\
2459949.71750 & 26.04 & 0.20 & 100.1 & 2460373.66213 & 23.98 & 0.20 & 66.3 \\
2459951.70374 & 25.66 & 0.19 & 92.0 & 2460375.66319 & 21.05 & 0.19 & 73.5 \\
2459952.65954 & 24.45 & 0.19 & 82.1 & 2460377.65147 & 17.56 & 0.19 & 55.9 \\
2459953.67722 & 22.41 & 0.24 & 81.3 & 2460378.60786 & 15.64 & 0.19 & 55.3 \\
2459955.75295 & 16.69 & 0.21 & 83.4 & 2460380.60788 & 11.59 & 0.24 & 54.1 \\
2459958.71240 & 8.33 & 0.23 & 69.7 & 2460382.60838 & 8.53 & 0.20 & 60.2 \\
2459960.64822 & 7.17 & 0.23 & 31.6 & 2460384.58757 & 9.48 & 0.27 & 20.2 \\
2459960.77232 & 7.13 & 0.20 & 82.9 & 2460385.60870 & 10.63 & 0.19 & 64.5 \\
2459962.68398 & 8.62 & 0.21 & 74.6 & 2460387.60896 & 15.31 & 0.20 & 65.3 \\
2459964.73495 & 12.63 & 0.20 & 90.2 & 2460390.60939 & 22.39 & 0.19 & 67.1 \\
2459966.69032 & 17.84 & 0.20 & 92.4 & 2460392.61010 & 24.57 & 0.18 & 68.5 \\
2459968.70462 & 22.91 & 0.19 & 93.4 & 2460394.58930 & 24.15 & 0.19 & 38.5 \\
2459970.66685 & 25.81 & 0.20 & 97.0 & 2460395.63619 & 22.94 & 0.20 & 52.6 \\
2459973.67019 & 24.59 & 0.18 & 33.9 & 2460397.63375 & 19.49 & 0.18 & 60.2 \\
2459974.64025 & 22.76 & 0.21 & 91.6 & 2460399.63827 & 15.60 & 0.25 & 60.5 \\
2459976.70143 & 17.21 & 0.21 & 69.7 &  &  &  &  \\
\noalign{\smallskip}\hline
\end{tabular}
\end{center}
\end{table*}

\end{document}